\documentclass[journal=jacsat,manuscript=article]{achemso}
\setkeys{acs}{articletitle = true} %to show titles in references
\usepackage{xr} 
\usepackage{longtable}
\usepackage{multirow}
\usepackage{amsmath}
\usepackage{amssymb}
\usepackage{subfigure}
\usepackage[usenames,dvipsnames]{color}
\SectionNumbersOn
\author{Markus Meuwly} \affiliation{Department of Chemistry,
  University of Basel, Klingelbergstrasse 80 , CH-4056 Basel,
  Switzerland.}  \email{m.meuwly@unibas.ch} \date{\today}

\title{Computational Vibrational Spectroscopy}

\begin{document}

\begin{abstract}
Vibrational spectroscopy is a powerful technique to characterize the
near-equilibrium dynamics of molecules in the gas- and the
condensed-phase. This contribution summarizes efforts from
computer-based methods to gain insight into the relationship between
structure and spectroscopic response. Methods for this purpose include
physics-based empirical energy functions, machine-learned force
fields, and methods that separate sampling conformational space and
determining the data for spectral analysis such as map-based
approaches.
  \end{abstract}

\section{Introduction}
Optical spectroscopy, and in particular infrared and Raman
spectroscopy, are versatile tools to determine the chemical
composition and structure of molecules. When extended into the
time-domain, techniques such as multidimensional infrared spectroscopy
also provide a structure-sensitive instrument to characterize the
environmental dynamics, couplings and energy transfer in solution. For
the interpretation of such experiments, simulations play an important
role. In the following, different approaches are summarized which aim
at providing a molecularly refined picture of the structural dynamics
underlying the experimentally observed spectroscopic features.\\

\noindent
The present contribution revolves around dynamics-based approaches
which explicitly account for the structural dynamics in
solution. Methods based on electronic structure calculations can
typically only be applied to individual conformations in the gas
phase. This has been done, for example, for small peptides for which
experimentally measured conformer-specific spectra are available. The
underlying structures were assigned by comparing normal modes
determined from density functional theory (DFT) calculations for
optimized structures sampled from either finite-temperature MD or
Monte Carlo simulations.\cite{zwier:2006,rizzo:2009}\\

\noindent
The central quantity for dynamics-based approaches to vibrational
spectroscopy is the potential energy surface (PES) which describes how
the total energy of a system changes with geometry. Computing and
representing a full-dimensional PES suitable for vibrational
spectroscopy is a formidable task in itself. Several methods available
for this are briefly mentioned and typical examples are
highlighted. The approaches range from augmented empirical energy
functions with physics-based input such a multipolar
electrostatics\cite{MM.rev:2020} to machine-learned energy
functions.\cite{unke:2021}

\section{Molecular Dynamics Simulations with Physics-Based Energy Functions}
Empirical energy functions - which are also called ``force fields'' -
have been extensively used to characterize the structure and dynamics
of macromolecules, including peptides and
proteins.\cite{lifson68,Levi69,Hagler942,Hagler941,charmm-md,amber,opls}
The extensive parametrization of any ``general purpose force field''
includes fitting to experimental structural and spectroscopic data for
equilibrium geometries and force constants, experimental results on
hydration free energies, heats of formation and other thermodynamic
properties for van der Waals parameters, and to electronic structure
data for atomic charges. As such, these energy functions are a useful
zeroth order model for a wide variety of problems in structural
biology and chemistry. However, for individual systems and specific
observables more refined parametrizations are required and possible.\\

\noindent
One particularly informative observable is the infrared spectrum of a
solvated molecule. The solvent-induced red or blue shifts of the
spectral lines provide a quantitative measure of the solvent-solute
interaction. If time-resolved methods such as 2-dimensional IR
spectroscopy are used, the frequency-fluctuation correlation function
is an observable and reflects the characteristic time-scale(s) of the
solvent fluctuations to which the solute degrees of freedom are
coupled.\cite{hamm2011concepts} These time scales are in the
picosecond range and at the core of the NCCR MUST. As such, they are
ideally suited for rigorous sampling by MD simulations and with
sufficiently long integration time the necessary averaging over
conformational substates can be achieved for direct comparison with
experiments. Typical time scales of such simulations are in the
nanosecond range which are sufficient for systems such as ions in
solution or ligands bound to proteins but not necessarily for ionic
liquids or deep eutectic solvent with considerably increased
viscosity.\\

\begin{figure}
 \begin{center}
 \resizebox{0.99\columnwidth}{!}
           {\includegraphics[scale=0.1,clip,angle=0]{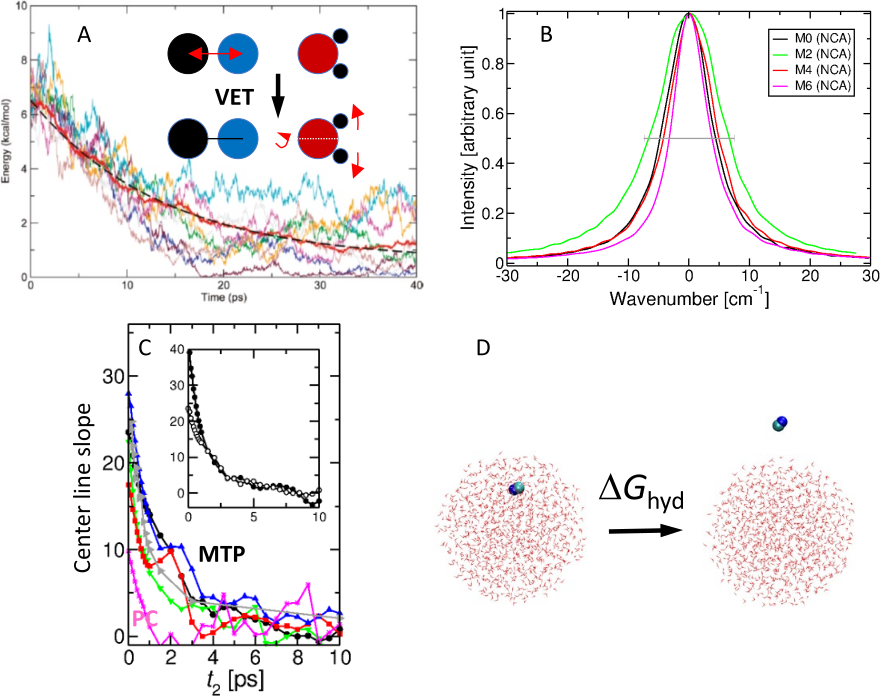}}
           \caption{Structural and hydration dynamics of cyanide in
             water. Panel A: vibrational relaxation of the $v_{\rm
               CN^-} = 1$ stretch through vibrational energy transfer
             (VET)\cite{MM.cn:2011} into the water libration and
             bending motion (red arrows) from classical MD simulations
             consistent with experiment.\cite{hamm:1997} Panel B: The
             1d lineshape from MTP simulations\cite{MM.cn:2013} with
             those measured experimentally (horizontal
             bar).\cite{hamm:1997} Panel C: Decay of the center line
             slope (equivalent to the FFCF) for PC and MTP charge
             models compared with experiment
             (grey).\cite{MM.cn:2013,hamm:2007} Panel D: Hydration
             free energy of CN$^{-}$ from MTP simulations are
             consistent with those from
             experiment.\cite{MM.cn2:2013,pearson:1986} Figure adapted
             from Refs.\cite{MM.cn:2013,MM.cn:2011,MM.cn2:2013}}
\label{fig:cn}
\end{center}
\end{figure}

\noindent
Atomistic simulations with multipolar force fields have demonstrated
that it is possible to realistically describe the 1d- and
2d-spectroscopy of small molecules in electrostatically demanding
environments such as in proteins or in
water.\cite{MM.mbco:2003,MM.mbco:2008,MM.cn:2013} It was also shown
for cyanide (CN$^-$) in water that the same energy function for the
solute is capable of correctly describing a range of condensed-phase
properties including the solvent-induced shift, the decay time of the
FFCF, the hydration free energy, and the vibrational energy relaxation
rate in water.\cite{MM.cn:2013,MM.cn:2011,MM.cn2:2013} This indicates
that physics-based refinements of such generic energy functions
provide a meaningful extension for molecularly resolved investigations
of complex systems.\\

\noindent
The spectroscopy of photodissociated CO in myoglobin (Mb) was also
investigated by using a fluctuating multipolar representation for the
electrostatics.\cite{MM.mbco:2003,MM.mbco:2008} This allowed to assign
for the first time the experimentally
observed\cite{alben:1982,lim:1995} split infrared spectrum and to
identify the two peaks with two distinct conformational substates: one
in which the oxygen end of CO pointed towards the heme-iron atom and a
second state for which the carbon was closer to the Fe
atom.\cite{MM.mbco:2003,MM.mbco:2008,MM.mbco:2006} This model
correctly captures the red shift of the two bands relative to gas
phase CO, the splitting of the two peaks and their relative intensity.\\

\begin{figure}
 \begin{center}
 \resizebox{0.99\columnwidth}{!}
           {\includegraphics[scale=0.1,clip,angle=0]{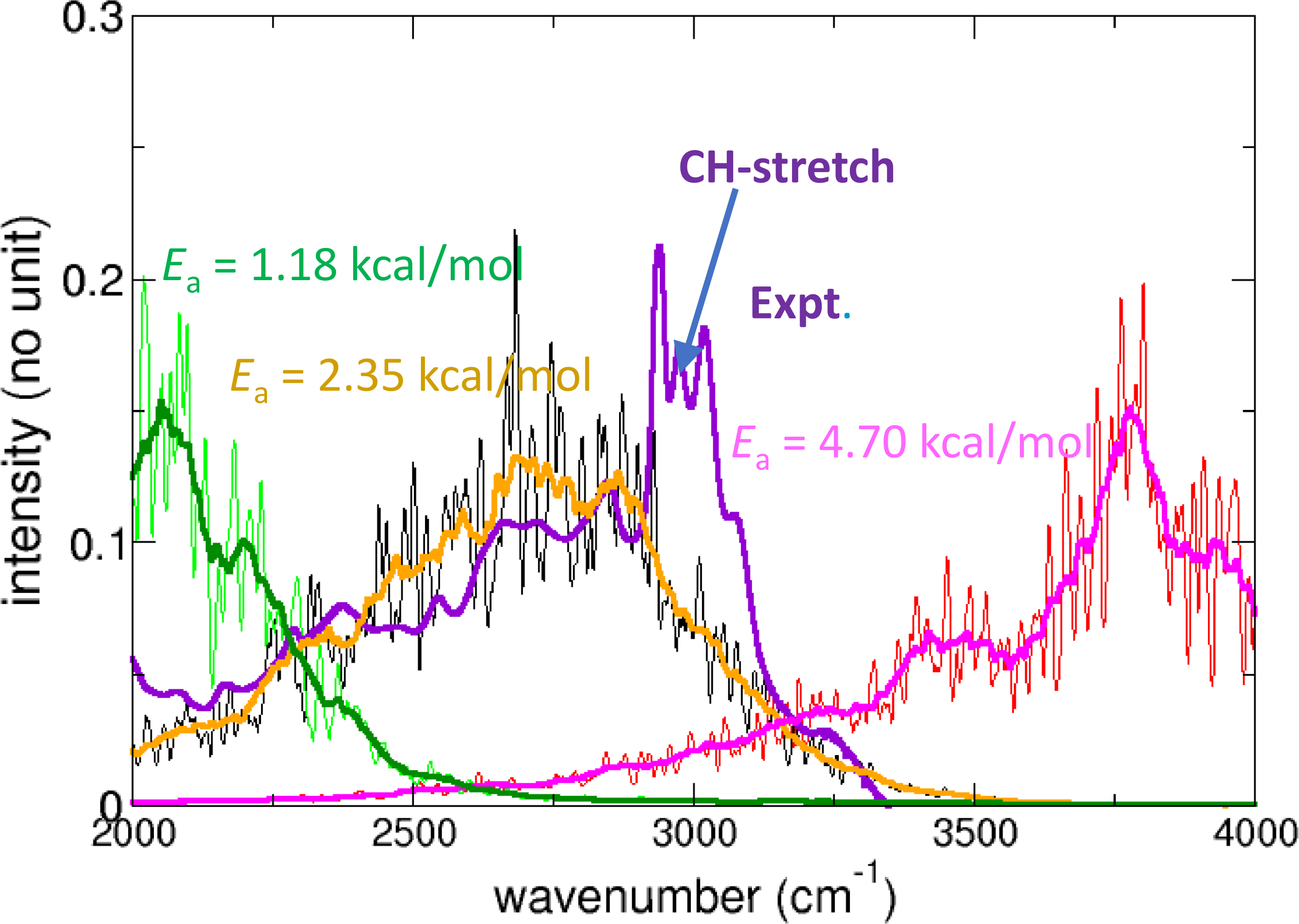}}
           \caption{Experimental infrared (blue) and computed (green,
             black/orange, red) power spectrum in the region of the
             H-transfer mode of acetylacetone. For the computations
             PESs featuring different barrier heights (1.18, 2.35,
             4.70) kcal/mol were used in a morphing-type
             approach\cite{MMmorphing99} to assess the position of the
             proton transfer band. Best agreement between
             experimentally measured and computed IR spectra is for a
             barrier height of 2.35 kcal/mol, compared with 3.2
             kcal/mol from CCSD(T) calculations.\cite{MM.acac:2015}
             The signatures around 3000 cm$^{-1}$ in the
             experimentally measured spectra are due to the CH stretch
             vibrations. Figure adapted from Ref.\cite{MM.acac:2015}}
\label{fig:acac}
\end{center}
\end{figure}

\noindent
Finally, it is also possible to refine energy functions by comparing
experimentally determined IR spectra with those from computations. The
infrared spectrum of acetylacetone (doubly methylated malonaldehyde)
features a prominent band between 2000 and 3000 cm$^{-1}$ which is due
to proton transfer across a low barrier.\cite{MM.acac:2015}
Morphing\cite{MMmorphing99,bowman:1991} a parametrized PES suitable
for following proton transfer and comparing the resulting infrared
spectrum with that from experiments yields an estimated barrier for
proton transfer of 2.35 kcal/mol; see Figure
\ref{fig:acac}. Subsequent machine learning found a barrier of 3.25
kcal/mol from transfer learning to the PNO-LCCSDT(T)-F12 level of
theory.\cite{MM.ht:2020}\\

\section{Molecular Dynamics Simulations with Machine-Learned Potentials}
One of the major challenges in empirical force field development is to
find a suitable parametrized form of the energy function depending on
internal coordinates for a molecule. Various extensions to the generic
harmonic oscillator models for bonds and angles have been
considered. But all of them incur a considerably larger number of free
parameters to be determined.\cite{mmff:1996,hagler:2015}\\

\noindent
In recent years an alternative approach has matured which is based on
using statistical models\cite{vapnik:1998} to represent precomputed
energies and forces from electronic structure calculations. Such
machine learning (ML) techniques do not necessarily require a
parametrized form to be used but rather represent data given a set of
kernel functions (kernel ridge regression) or by minimizing a loss
function of a neural network
(NN).\cite{manzhos:2020,unke:2021,MM:2021} For kernel-based methods
the long range physical shape of the PES can be encoded in the kernel
which guarantees correct extrapolation to large
separations.\cite{ho96:2584,unk17:1923} No such procedure is known for
short range interactions.\cite{soldan:2000} For NN-learned energy
functions extrapolation to geometries outside the training set needs
to be carefully assessed.\cite{unke:2021}\\

\noindent
One of the advantages of ML-based energy functions is that they
contain all couplings between the degrees of freedom. This is very
challenging for empirical energy functions. For instance, the CO bonds
in protonated oxalate change between single- and double-bond character
depending on where the proton is located.\cite{MM.oxa:2017} Although
such effects can be ``encoded'' in an empirical force field, capturing
such effects from a globally valid, machine-learned energy function is
more readily possible as has recently been done for formic acid
dimer.\cite{MM.tl:2022,MM.fad.2:2022}\\

\noindent
As an example for the performance of state-of-the art ML-based methods
for vibrational spectroscopy, formic acid monomer and dimer (FAM and
FAD) in the gas phase is considered.\cite{MM.fad:2022} Using
PhysNet\cite{MM.physnet:2019} a reference machine-learned PES was
determined at the MP2/aug-cc-pVTZ level of theory for FAM and FAD. The
mean averaged error between reference calculations and the statistical
model is 0.01 kcal/mol. Transfer learning the MP2-based PES to the
CCSD(T)/aug-cc-pVTZ level of theory yields normal mode frequencies
within 25 cm$^{-1}$ on average compared with experiment for modes
below 2000 cm$^{-1}$. Including anharmonic corrections within second
order vibrational perturbation theory
(VPT2)\cite{barone2005anharmonic} reduces this to 17 cm$^{-1}$. For
the OH-stretch mode the VPT2 calculations yield 3011 cm$^{-1}$,
compared with an experimentally reported, $\sim 100$ cm$^{-1}$ broad
absorption band with center frequency at $\sim 3050$
cm$^{-1}$. Finally, using diffusion Monte Carlo
(DMC)\cite{anderson1975random} calculations for the full-dimensional
ground state potential energy and including corrections due to basis
set superposition and basis set completeness errors yield a
dissociation energy of $D_0 = -14.23 \pm 0.08$ kcal/mol compared with
an experimentally determined value of $-14.22 \pm
0.12$~kcal/mol.\cite{suhm:2012}\\

\noindent
It is of interest to note that experiment-guided refinement of an
advanced force field based on molecular mechanics with proton transfer
(MMPT)\cite{MM.mmpt:2008} the barrier height for double proton
transfer in FAD could be inferred. For this, the height of the double
well potential was adjusted to match the experimentally observed broad
band associated with DPT in FAD. The resulting\cite{MM.fad:2016}
barrier height was 7.2 kcal/mol which compares with an experimentally
determined value from microwave spectroscopy of 7.3
kcal/mol.\cite{caminati:2019} In this fashion, information from
vibrational spectroscopy can also be used to adapt (``morph'')
PESs.\cite{MMmorphing99}\\

\noindent
Finally, machine-learned energy functions can also be used in a mixed
quantum mechanics/molecular mechanics fashion to accurately describe
the bonded energetics for spectroscopic probes used in protein
2-dimensional IR spectroscopy.\cite{johnson2017quantifying} This was
successfully done using reproducing kernel-based representations for
the amide-I mode in insulin and trialanine or for azide attached to
all alanine residues in
Lysozyme.\cite{MM.insulin:2020,MM.lys:2021,MM.ala3:2021} This provides
a positionally sensitive probe of the protein dynamics to follow
protein assembly or protein-ligand interactions.\\

\section{Molecular Dynamics Simulations with Spectroscopic Maps}
Determining the frequency trajectories $\omega(t)$ required for
computing the FFCF and 2d-IR response can be computationally
prohibitive. One way to circumvent the expensive instantaneous normal
mode or reduced-dimensionality quantum bound state calculations is to
use spectroscopic
maps.\cite{knoester-jpc-model-2006,Tokmakoff-map-jcp-2013,skinner-map-JPCB2011,baiz:2020}
Such maps can be parametrized from electronic structure calculations
using model systems and exploit the fact that the frequency shift of
an oscillator in the field of surrounding point charges can be
approximately described by the Stark effect. Maps have been generated
for a range of spectroscopic probes, including the amide-I
stretch,\cite{Tokmakoff-map-jcp-2013}, the nitrile stretch, the azido
stretch, and others.\cite{baiz:2020} Recently, machine learning has
been applied to refine the amide-I map.\cite{skinner:2019}\\

\noindent
With such spectroscopic maps it is then quite straightforward to
determine the frequency trajectory for a particular oscillator from
conventional MD simulations. For every snapshot to be analyzed the
electric field at the position of the oscillator of interest is
determined and related to the frequency shift by evaluating the
spectroscopic map. This provides the information required to generate
the FFCF from which important information about the structural
dynamics around the spectroscopic reporter can be obtained.\\

\noindent
One of the conceptual disadvantages of spectroscopic maps is the fact
that the energy function used to run the MD simulations typically
differs from the energy function used to evaluate the spectroscopic
response. This can be done from physics-based force
fields. Furthermore, some maps have been generated for rigid labels as
was the case for the amide-I maps. Hence, the MD simulations need to
be run with constrained CO distances for the maps to be valid. A
direct comparison between map-based analyses and results from
instantaneous normal modes and solutions of the nuclear Schr\"odinger
equation has been recently been given for insulin monomer and
dimer. For this system it was found that the maps perform inferior
compared with the other two approaches.\cite{MM.insulin:2020}\\

\section{Outlook}
Molecular dynamics simulations with advanced energy functions provide
important information about the structural dynamics of molecules in
solution. Extensively sampling the conformational degrees of freedom
is essential and only possible with efficient implementations of the
energy functions. As with the spectroscopic probes to characterize
local protein dynamics simulations based on quantitatively accurate
energy functions provide information where to insert such probes in
order to be most sensitive to external perturbations such as binding
of a ligand. For such and other applications the combination of
experiment and simulation is indispensable and promises the necessary
molecular-level information to control and design chemical systems
with desired properties.\\

\noindent
For the nuclear dynamics classical MD simulations have proven adequate
for the purposes outlined in the present contribution. It is of
interest to note that already 40 years ago it was pointed out for CO
in the gas phase and in argon classical MD simulations are even
capable of capturing R- and P-branches similar to what quantum
mechanical treatments provide.\cite{wilson:1981} More recently, this
has also been found for liquid water.\cite{skinner:2019} Hence, it is
anticipated that such MD simulations with improved energy functions
can contribute even more to understanding of the vibrational
spectroscopy of solvated species than was previously
anticiptated. Nevertheless, approximate quantum treatments will be of
interest in order to delineate the range of applicability of classical
mechanics for vibrational spectroscopy. Also, quantum effects
including zero-point energy and tunneling are outside the scope of any
classical mechanics-based method. For this reason, developing quantum
methods applicable to the dynamics in solution remain an important
quest in this field.\cite{rossi:2021}

\section*{Acknowledgments}
The author acknowledges financial support from the Swiss National
Science Foundation (NCCR-MUST and Grant No. 200021-188724), the AFOSR,
and the University of Basel.

\bibliography{refs.tidy}

\end{document}